\documentclass[conference]{IEEEtran}
\usepackage[normalem]{ulem}
\usepackage{color}

\usepackage{amsmath}
\usepackage{mathrsfs}

\ifCLASSINFOpdf
\usepackage[pdftex]{graphicx}
\else
\fi
\usepackage{algorithm}
\usepackage{algpseudocode}

\begin{document}

%
\title{Passive Multi-Target Tracking\\ Using the Adaptive Birth Intensity PHD Filter}

\author{
\IEEEauthorblockN{Christopher Berry\IEEEauthorrefmark{1}, Donald J. Bucci\IEEEauthorrefmark{1}, Samuel Watt Schmidt\IEEEauthorrefmark{1}}
    \IEEEauthorblockA{\IEEEauthorrefmark{1}Lockheed Martin Advanced Technology Laboratories, Cherry Hill, NJ, USA
    \\Email: \{christopher.m.berry, donald.j.bucci.jr, samuel.w.schmidt\}@lmco.com}
}

\maketitle


\begin{abstract}
Passive multi-target tracking applications require the integration of multiple spatially distributed sensor measurements to distinguish true tracks from ghost tracks.
A popular multi-target tracking approach for these applications is the particle filter implementation of Mahler's probability hypothesis density (PHD) filter, which jointly updates the union of all target state space estimates without requiring computationally complex measurement-to-track data association. Although this technique is attractive for implementation in computationally limited platforms, the performance benefits can be significantly overshadowed by inefficient sampling of the target birth particles over the region of interest.
We propose a multi-sensor extension of the adaptive birth intensity PHD filter described in (Ristic, 2012) to achieve efficient birth particle sampling driven by online sensor measurements from multiple sensors.
The proposed approach is demonstrated using distributed time-difference-of-arrival (TDOA) and frequency-difference-of-arrival (FDOA) measurements, in which we describe exact techniques for sampling from the target state space conditioned on the observations. Numerical results are presented that demonstrate the increased particle density efficiency of the proposed approach over a uniform birth particle sampler.
\end{abstract}


%
\IEEEpeerreviewmaketitle


\section{Introduction}
Passive localization and tracking of multiple co-channel targets is an important research topic in speaker localization \cite{Ma2006b}, sonar \cite{Zhang2017}, passive radar \cite{Tian2006}, and intrusion detection in wireless sensor networks \cite{Vempaty2014}. By \emph{passive}, we mean that the signal shape emitted by the target is unknown, but observed similarly across multiple spatially distributed sensors. As a result, the localization and tracking algorithms for these applications rely on \emph{difference} measurements between pairs of sensors. Time-difference-of-arrival (TDOA) and frequency-difference-of-arrival (FDOA) measurements, for example, use the generalized cross correlation of raw signal measurements between sensor pairs at different time and frequency offsets to estimate range and range-rate differences \cite{Vankayalapati2012,Holt1987}. Power-difference-of-arrival (PDOA) measurements are generated by observing the difference in average receive power measurements between sensors, which is closely related to the difference in the cross correlation peak magnitudes. In applications such as underwater acoustics and passive radar a direction finding solution based on angle-of-arrival (AOA) is also possible \cite{Zhang2017}.

Although single-target localization and tracking algorithms using passive measurements are well understood \cite{Vankayalapati2014}, the problem becomes significantly more complex for multiple co-channel targets. For passive localization and tracking algorithms in particular, the presence of multiple targets leads to state space estimation ambiguities known as \emph{ghost targets}. These ghost targets manifest from the intersection of plausible state space estimates suggested by each passive sensor measurement \cite{ElGonnouni2014}. Resolution of these ghost targets requires specification of \emph{measurement-to-track association} algorithms that in general do not scale well with the number of targets, especially in the presence of high clutter and high false alarm rate environments.

A number of multi-target tracking approaches have been proposed for the passive multi-target tracking problem based on the random finite sets (RFS) framework \cite{Guldogan2012,Ozkan2011}. These techniques extend the calculus and statistics used to derive the standard recursive Bayes filter to the case where the cardinality of the underlying state space is itself a random variable. The resulting \emph{multi-target recursive Bayes filter} provides a mechanism for estimating the number of targets and their state space estimates without having to specify a measurement-to-track association technique. 
Although this filter is practically difficult to implement, a number of moment-based approximations have been proposed that are widely used in computationally limited settings.
The \emph{probability hypothesis density (PHD)} filter results from a first-order moment approximation where the cardinality distribution is assumed Poisson.
The \emph{cardinalized probability hypothesis density (CPHD)} filter results from a first-order moment approximation where the cardinality distribution is estimated in parallel.
Both of these approximation techniques are defined for single sensor tracking problems, as the original multi-sensor update equation requires the combinatorial evaluation of all false-alarm and missed detection hypotheses per sensor. In practice, the \emph{iterated-corrector PHD/CPHD} method is used which involves applying the update step sequentially to the set of measurements from each sensor \cite{Meng2009,Tian2006}. Although this approach is suboptimal \cite{Mahler2010,Zhang2010}, it remains practical in many applications due to the marginal increase in complexity.

In this paper, we focus on the particle filter implementation of the iterated-corrector PHD filter \cite{Vo2003,Meng2009,Tian2006} applied to the passive multi-target tracking problem using a stationary wireless sensor network.
A set of spatially distributed sensors observe co-channel, non-overlapping (in time) signals from a time varying number of targets. Each sensor observes pairs of TDOA and FDOA measurements that are generated from at most one of the targets, or the result of environmental clutter. The goal is to use the measurements at each time step to track the position and velocity of all targets.
In this work, we accomplish this task using the \emph{adaptive birth intensity} extension to the  PHD filter detailed in \cite{Ristic2012a}.
This approach constructs an augmented state space consisting of the original state-space variables concatenated with a label that distinguishes a newborn target from a persistent target. The resulting predict and update steps of the PHD filter after applying this augmentation removes the need for prior specification of the target birth intensity, and instead replaces it with a particle sampling procedure based on the measurements themselves.
More importantly, this data-adaptive sampling technique enables very accurate multi-target state estimation without requiring an excessive number of particles.
This consideration is important, especially when considering implementation on computationally limited software defined radio (SDR) platforms.

 In order to implement this technique for the TDOA/FDOA passive localization problem, we provide the following contributions. First, we describe an extension of the adaptive birth intensity technique provided in \cite{Ristic2012a} to multiple sensor measurements using the iterated-corrector implementation of the particle PHD filter.
 The approach of \cite{Ristic2012a} provides a more rigorous justification for an adaptive birth intensity, in contrast to similar heuristics previously suggested in \cite{Lanterman2008,XingbaoWang2016}.
 As opposed to the linear Gaussian approximation of the birth intensity sampler suggested in \cite{Ristic2012a}, we also provide a mechanism for directly sampling the plausible state-space particles using the TDOA and FDOA measurement tuples. Our technique is based on the geometric models presented previously in \cite{Musicki2008}.
 The specification of these sampling functions is very important, as the relationship between the state space estimates and the TDOA and FDOA measurements are non-linear and non-invertible.

The remainder of the paper is organized as follows. In Section~\ref{sec:rfs_prelim}, a brief overview of RFS for multi-target tracking is provided. The adaptive birth intensity particle PHD filter from \cite{Ristic2012a} alongside our extension to multi-sensor measurements using the iterated-corrector update rule is summarized in Section~\ref{sec:ic_abif}. We then detail the TDOA and FDOA measurement models alongside the relevant position and velocity sampling techniques in Section~\ref{sec:sampling}. Finally, Section~\ref{sec:simulation} provides a simulation comparing the performance of the proposed technique with a standard uniform placement of birth particles.


\section{Random Finite Set Preliminaries}\label{sec:rfs_prelim}

A full treatment of the multi-target tracking problem with RFS is outside the scope of this paper.
Instead, we provide a pragmatic description using the notation provided in \cite{Ristic2012a}.
A detailed description of RFS can be found in \cite{Mahler2007}, and its application to the particle PHD filter in \cite{Vo2003}.


\subsection{Random Finite Sets and Multi-target Tracking}
Consider a multi-target tracking problem where the goal is to estimate the state vector, $\mathbf{x}_k \in \mathscr{X}$, for $N_k$ targets at the discrete time step $k = 0,1,\ldots$. At each time step, the number of targets is random, and the states for each target evolve according to a Markov process. The \emph{multi-target state} at time $k$ is defined as
$$
\mathbf{X}_k = \{\mathbf{x}_{k,1},\ldots,\mathbf{x}_{k,n_k}\} \in \mathscr{F}(\mathscr{X})
$$
where $\mathscr{F}(\mathscr{X})$ denotes a finite collection of subsets defined on $\mathscr{X}$. A single sensor generates a set of measurements, $\mathbf{z}_k \in \mathscr{Z}$, for each target in addition to false alarm measurements (\textit{i.e.}, clutter) that are uncorrelated with the any of the target states. In addition, missed detections may occur in which no measurements are observed from a given target. The total set of $m_k$ measurements at time step $k$, consisting of true target and clutter returns, is defined as
$$
\mathbf{Z}_k = \{\mathbf{z}_{k,1},\ldots,\mathbf{z}_{k,m_k}\} \in \mathscr{F}(\mathscr{Z})
$$
Both $\mathbf{X}_k$ and $\mathbf{Z}_k$ are described as \emph{random finite sets} (RFS), where the values of the elements and the cardinality of the sets are random variables with no specific element permutation (\textit{i.e.}, ordering of the targets). 
In a manner similar to single-target filtering, a RFS multi-target filter seeks to estimate the posterior multi-target state distribution, $f_{k|k}(\mathbf{X}_k|\mathbf{Z}_{1:k})$, which relates the multi-target state estimates $\mathbf{X}_k$ given the measurements up until time $k$, denoted $\mathbf{Z}_{1:k}=\{\mathbf{Z}_1,\ldots,\mathbf{Z}_k\}$.
This RFS posterior is fully specified using the multi-target recursive Bayes filter derived in \cite{Mahler2007}.


\subsection{Probability Hypothesis Density Filtering}
Although particle filter implementations of the multi-target recursive Bayes filter exist \cite{Ma2006b}, they are not often used in practice because their complexity becomes intractable for large numbers of targets.
Instead, a number of RFS moment approximations are applied to arrive at a suboptimal but more tractable realization.
The moment approximation of interest for this work is the \emph{probability hypothesis density (PHD)}.
The PHD, denoted $D(\mathbf{x}):\mathscr{X}\to\Re$, defines a multi-target analogue to an expected value in the sense that it provides a point-wise estimate of the density of targets located at $\mathbf{x}_k \in \mathscr{X}$ \cite{Ristic2012a,Mahler2007}.
As a consequence, the standard integral of the PHD function across the state space $\mathscr{X}$ evaluates to the expected number of targets in the state space. 

In a multi-target tracking system the PHD function can be loosely interpreted as an estimate of the union (\textit{i.e.}, weighted summation) of single-target tracker posteriors and a prior density function consisting of a birth probability distribution over the state space multiplied by the expected number of newborn targets.
Denote the most recent estimate of the PHD intensity function at time $k-1$ as $D_{k-1|k-1}(\mathbf{x})$. The PHD filter \emph{prediction step} is given by
\begin{multline}\label{eq:phd_pred}
D_{k|k-1}(\mathbf{x}) = \gamma_{k|k-1}(\mathbf{x}) \\
+ p_s(\mathbf{x})\int \pi_{k|k-1}(\mathbf{x}|\mathbf{x}')D_{k-1|k-1}(\mathbf{x})d\mathbf{x}'
\end{multline}
where $p_s(\mathbf{x})$ is the probability of survival for a target, $\pi_{k|k-1}(\mathbf{x}|\mathbf{x}')$ is the standard state-transition probability density function that describes how target kinematics propagate from time step $k-1$ to $k$, and $\gamma_{k|k-1}(\mathbf{x})$  is a PHD function representing the birth process (\textit{i.e.}, the expected number of targets that appear at each state space location). The PHD filter \emph{update step} is given by
\begin{multline}\label{eq:phd_upd}
D_{k|k}(\mathbf{x}) = \left(1-p_d(\mathbf{x})\right)D_{k|k-1}(\mathbf{x}) \\
+\sum_{\mathbf{z}_k \in \mathbf{Z}_k} \frac{g_{k}(\mathbf{z}_k|\mathbf{x}) p_d(\mathbf{x})}{\kappa_k(\mathbf{z}_k) + C(\mathbf{z}_k)}D_{k|k-1}(\mathbf{x})
\end{multline}
where $p_d(\mathbf{x})$ is the probability of detection of a measurement being generated from a target, $g_{k}(\mathbf{z}_k|\mathbf{x})$ is the standard single-target measurement likelihood, $\kappa_k(\mathbf{z}_k)$ is a PHD function representing the measurement clutter process (\textit{i.e.}, the expected number of measurement false alarms that appear at each measurement space location).
The normalization term $C(\mathbf{z}_k)$ is given as
\begin{equation}
C(\mathbf{z}_k) = \int g_{k}(\mathbf{z}_k|\mathbf{x}')p_d(\mathbf{x}')D_{k|k-1}(\mathbf{x}')d\mathbf{x}'.
\end{equation}
%
Since the PHD filter involves multiple standard integrals over the state space, it is typically implemented using a sequential Monte Carlo (\textit{i.e.}, particle filter) realization \cite{Vo2003}, or using Gaussian mixture distributions \cite{Vo2006}. When multiple sensors are present, the iterated-corrector approach applies the update step of Equation~\eqref{eq:phd_upd} sequentially for the set of measurements obtained at each sensor.


\section{Multi-sensor Adaptive Birth Intensity Filter}\label{sec:ic_abif}

\subsection{Adaptive Birth Intensity PHD Filter}\label{sec:abi_phd}
The adaptive birth intensity technique of \cite{Ristic2012a} augments the original state space by appending a label, $\beta\in\{0,1\}$, which distinguishes a newborn target ($\beta=1$) from a persisting target ($\beta = 0$). 
Augmenting the original state vector of the PHD predict and update equations with this label decouples the birth PHD intensity from the predicted PHD intensity of persisting targets.
The PHD filter predict step for this augmented state space, denoted $D_{k|k-1}(\mathbf{x},\beta)$ is given in \cite{Ristic2012a} as
\begin{multline}\label{eq:abi_phdpred_beta0}
D_{k|k-1}(\mathbf{x},0) =\\
p_s(\mathbf{x})\int \pi_{k|k-1}(\mathbf{x}|\mathbf{x}')\left[D_{k-1|k-1}(\mathbf{x},1) \right.\\\left.+ D_{k-1|k-1}(\mathbf{x},0)\right]d\mathbf{x}'
\end{multline}
and
\begin{equation}\label{eq:abi_phdpred_beta1}
D_{k|k-1}(\mathbf{x},1) = \gamma_{k|k-1}(\mathbf{x}).
\end{equation}
%
The PHD update step for this augmented state space is
\begin{multline}
D_{k|k}(\mathbf{x},0) =\left(1-p_d(\mathbf{x})\right)D_{k|k-1}(\mathbf{x},0) \\
+\sum_{\mathbf{z}_k \in \mathbf{Z}_k} \frac{g_{k}(\mathbf{z}_k|\mathbf{x}) p_d(\mathbf{x})D_{k|k-1}(\mathbf{x},0)}{\kappa_k(\mathbf{z}_k) + C'(\mathbf{z}_k)}
\end{multline}
and
\begin{equation}
D_{k|k}(\mathbf{x},1) =\sum_{\mathbf{z}_k \in \mathbf{Z}_k} \frac{g_{k}(\mathbf{z}_k|\mathbf{x})\gamma_{k|k-1}(\mathbf{x})}{\kappa_k(\mathbf{z}_k) + C'(\mathbf{z}_k)}
\end{equation}
where the new normalization term $C'(\mathbf{z}_k)$ is given as
\begin{multline}
C'(\mathbf{z}_k) = \int g_{k}(\mathbf{z}_k|\mathbf{x}')\gamma_{k|k-1}(\mathbf{x}')d\mathbf{x}' \\+ \int g_{k}(\mathbf{z}_k|\mathbf{x}')p_d(\mathbf{x}')D_{k|k-1}(\mathbf{x}',0)d\mathbf{x}'.
\end{multline}
It is important to note that the addition of these target labels do not change the mechanics of the PHD filter, but instead provide insight as to how the birth PHD function affects the predict and update steps.
Newborn targets from the birth process are brought into the updated PHD function based on their product with the measurement likelihood function.
As discussed in \cite{Ristic2012a}, this insight allows for construction of a particle filtering solution where the birth intensity is set such that the particle density is high at areas of high measurement likelihood (\textit{i.e.}, $\int g_{k}(\mathbf{z}_k|\mathbf{x}')\gamma_{k|k-1}(\mathbf{x}')d\mathbf{x}'$ is large).

The particle filter implementation of the adaptive birth intensity PHD filter for a single sensor is specified by constructing two particle systems: one for persisting targets and one for newborn targets. These approximations are expressed as
$$
D_{k|k}(\mathbf{x},0) \approx \sum_{i=1}^{N^p_k}w_{k,p}^{(i)}\delta_{\mathbf{x}_{k,p}^{(i)}}(\mathbf{x})
$$
$$
D_{k|k}(\mathbf{x},1) \approx \sum_{i=1}^{N^b_k}w_{k,b}^{(i)}\delta_{\mathbf{x}_{k,b}^{(i)}}(\mathbf{x})
$$
where the sets $\cup_{i=1}^{N^p_k}(\mathbf{x}_{k,p}^{(i)},w_{k,p}^{(i)})$ and $\cup_{i=1}^{N^b_k}(\mathbf{x}_{k,b}^{(i)},w_{k,b}^{(i)})$ represent the particle systems for the persistent and newborn targets respectively. The Dirac delta function $\delta_{\mathbf{x}'}(\mathbf{x}) = 1$ when $\mathbf{x} = \mathbf{x}'$ and 0 otherwise. An additional particle system is defined to represent the sum of the persistent and newborn PHD intensity functions used in the predict step of Equation~\eqref{eq:abi_phdpred_beta0},
$$
D_{k|k}(\mathbf{x},0) + D_{k|k}(\mathbf{x},1) \approx \sum_{i=1}^{N_k}w_{k}^{(i)}\delta_{\mathbf{x}_{k}^{(i)}}(\mathbf{x})
$$
which is defined as the union of the persistent and newborn particle systems,
\begin{multline*}
\cup_{i=1}^{N_k}(\mathbf{x}_{k}^{(i)},w_{k}^{(i)}) =\\
 \left[\cup_{i=1}^{N^p_k}(\mathbf{x}_{k,p}^{(i)},w_{k,p}^{(i)})\right] \bigcup \left[\cup_{i=1}^{N^b_k}(\mathbf{x}_{k,b}^{(i)},w_{k,b}^{(i)})\right]
\end{multline*}
where $N_k = N^b_k+N^p_k$. 

With these three particle system approximations, the particle filter variation of the \emph{single target adaptive birth intensity PHD filter} is given as follows \cite{Ristic2012a}.
In the predict step, the persisting particle weights are determined by sampling all $\mathbf{x}_{k,p}^{(i)}$ from an importance density function $q_{k}(\mathbf{x}^{(i)}_{k,p}|\mathbf{x}^{(i)}_{k-1},\mathbf{Z}_k)$ and setting the weights according to
\begin{equation}\label{eq:abi_phdpred_persistweights}
w_{k|k-1,p}^{(i)} = p_s(\mathbf{x}_{k,p}^{(i)})\frac{\pi_{k|k-1}(\mathbf{x}^{(i)}_{k,p}|\mathbf{x}^{(i)}_{k-1})}{q_{k}(\mathbf{x}^{(i)}_{k,p}|\mathbf{x}^{(i)}_{k-1},\mathbf{Z}_k)}w_{k-1}^{(i)}.
\end{equation}
The newborn particles and their weights are determined by sampling $\mathbf{x}_{k+1,b}^{(i)}$ from some measurement conditioned probability distribution function $b(\mathbf{x}_{k+1,b}|\mathbf{z}_k)$ for each $\mathbf{z}_k \in \mathbf{Z}_k$. The probability density function $b(\mathbf{x}_{k+1,b}|\mathbf{z}_k)$ is selected such that all particles generated are in regions of space where the likelihood $g_{k+1}(\mathbf{z}_k|\mathbf{x}_{k+1,b})$ is high. The corresponding weights for the newborn particles are uniform, and given as
\begin{equation}\label{eq:abi_phdpred_birthweight}
w_{k|k-1,b}^{(i)} = \frac{\nu^b_{k|k-1}}{N^b_{k}}
\end{equation}
where $\nu^b_{k|k-1}$ is the expected number of target births at each predict step. In the update step, the predicted persistent particle weights are refined as
\begin{multline}\label{eq:abi_phdpred_persupd}
w^{(i)}_{k|k,p} = (1 - p_d(\mathbf{x}^{(i)}_{k|k-1,p}))w^{(i)}_{k|k-1,p}\\
+\sum_{\mathbf{z}\in\mathbf{Z}_{k}}\frac{p_d(\mathbf{x}^{(i)}_{k|k-1,p})g_{k}(\mathbf{z}|\mathbf{x}^{(i)}_{k|k-1,p})w^{(i)}_{k|k-1,p}}{\mathcal{L}(\mathbf{z})}
\end{multline}
where the normalization constant is given as
\begin{multline}
\mathcal{L}(\mathbf{z})=\kappa_k(\mathbf{z}) + \sum_{i=1}^{N_k^b}w_{k|k-1,b}^{(i)} \\
+\sum_{i=1}^{N_{k-1}}p_d(\mathbf{x}^{(i)}_{k|k-1,p})g_k(\mathbf{z}|\mathbf{x}^{(i)}_{k|k-1,p})w^{(i)}_{k|k-1,p}.
\end{multline}
Similarly, the birth particle weights are refined from the predicted birth particle set according to
\begin{equation}\label{eq:abi_phdpred_birthupd}
w^{(i)}_{k|k,b} = \sum_{\mathbf{z}\in\mathbf{Z}_{k}}\frac{w^{(i)}_{k|k-1,b}}{\mathcal{L}(\mathbf{z})}
\end{equation}
Once the update step is finished, a resampling step is applied to the persistent\footnote{As \cite{Ristic2012a} suggests, the resampling step can also be applied for the updated newborn particles, if necessary. In this work, we apply resampling only to the updated persistent particle systems.} PHD particle systems in order to remove samples with low weights. The usual approach is to compute the weight sum of each system, normalize the particle weights by each system's weight sum, resample the points using each particle system's normalized weights, and finally set the new particle weights uniformly such that they sum to the original weight sums of each system. A more detailed description is provided in \cite{Vo2003}.


\begin{figure*}[t!]
\renewcommand{\algorithmicrequire}{\textbf{Input:}}
\renewcommand{\algorithmicensure}{\textbf{Return:}}
\hrulefill\\
\textbf{Algorithm 1:} Multi-sensor adaptive birth intensity PHD filter at time step $k$\vspace{-.75em}

\hrulefill
\begin{algorithmic}[1]
\Require
	\Statex (1) Combined persistent and newborn PHD particle system from step $k-1$, $\cup_{i=1}^{N_{k-1}}(\mathbf{x}_{k-1}^{(i)},w_{k-1}^{(i)})$
	\Statex (2) Observed measurement set at step $k$ for each sensor $l$, $\cup_{l=1}^L\mathbf{Z}^{(l)}_k$ where $\mathbf{Z}^{(l)}_k = \{\mathbf{z}^{(l)}_{k,1},\ldots,\mathbf{z}^{(l)}_{k,m^{(l)}_k}\}$
\State \(\triangleright\) Sample all particles according to target state-transition dynamics
\For{$i = 1,\ldots,N_{k-1}$}
	\State Sample $\mathbf{x}^{(i)}_{k|k-1,p}$ from $q_{k}(\mathbf{x}^{(i)}_{k|k-1,p}|\mathbf{x}^{(i)}_{k-1},\mathbf{Z}_k)$
	\State $w_{k|k-1,p}^{(i)} \gets p_s(\mathbf{x}^{(i)}_{k|k-1,p})\frac{\pi_{k|k-1}(\mathbf{x}^{(i)}_{k|k-1,p}|\mathbf{x}^{(i)}_{k-1})}{q_{k}(\mathbf{x}^{(i)}_{k|k-1,p}|\mathbf{x}^{(i)}_{k-1},\mathbf{Z}_k)}w_{k-1}^{(i)}$ \Comment{Equation~\eqref{eq:abi_phdpred_persistweights}}
\EndFor 
\State \(\triangleright\) Iterated-corrector applied to each sensor measurement set sequentially
\For {$l=1,\ldots,L$}
	\State $N_k^b \gets M_b\cdot m^{(l)}_k$ \Comment{Sample $M_b$ particles per newborn target based on the number of measurements at sensor $l$}
	\For {$i=1,\ldots,m^{(l)}_k$}
		\For {$j=1,\ldots,M_b$}
			\State $n \gets j + (i-1)M_b$
			\State Sample $\mathbf{x}_{k|k-1,b}^{(n)} \sim b(\mathbf{x}_{k|k-1,b}^{(n)}|\mathbf{z}^{(l)}_{k,i})$ \Comment{Measurement-driven birth intensity sampling}
			\State $w_{k|k-1,b}^{(n)} \gets \frac{\nu^b_{k|k-1}}{N^b_{k}}$ \Comment{Equation~\eqref{eq:abi_phdpred_birthweight}}
		\EndFor
	\EndFor 	
	\State \(\triangleright\) Update step for persistent and newborn particles at step $k$
	\State Evaluate each $w_{k|k,p}^{(i)}$ using $\cup_{i=1}^{N_{k-1}}(\mathbf{x}_{k|k-1,p}^{(i)},w_{k|k-1,p}^{(i)})$, $\cup_{i=1}^{N_{k}^b}(\mathbf{x}_{k|k-1,b}^{(i)},w_{k|k-1,b}^{(i)})$, $\mathbf{Z}^{(l)}_k$, and Equation~\eqref{eq:abi_phdpred_persupd}
	\State Evaluate each $w_{k|k,b}^{(i)}$ using $\cup_{i=1}^{N_{k-1}}(\mathbf{x}_{k|k-1,p}^{(i)},w_{k|k-1,p}^{(i)})$, $\cup_{i=1}^{N_{k}^b}(\mathbf{x}_{k|k-1,b}^{(i)},w_{k|k-1,b}^{(i)})$, and Equation~\eqref{eq:abi_phdpred_birthupd}

	\If {$l < L$}
		\State \(\triangleright\) Combine updated persistent and newborn particle systems for the next sensor's update step
		\State $\cup_{i=1}^{ N_{k-1} + N_k^b}(\mathbf{x}_{k|k-1,p}^{(i)},w_{k|k-1,p}^{(i)}) \gets \left[\cup_{i=1}^{N_{k-1}}(\mathbf{x}_{k|k-1,p}^{(i)},w_{k|k,p}^{(i)})\right] \bigcup \left[\cup_{i=1}^{N_{k}^b}(\mathbf{x}_{k|k-1,b}^{(i)},w_{k|k,b}^{(i)})\right]$
	\EndIf
\EndFor
\State \(\triangleright\) Resample persisting PHD particle system after the final sensor update
\State $\hat{\nu}_k^p \gets \sum_{i=1}^{N_{k-1}}w_{k|k,p}^{(i)}$ \Comment{Estimated number of persisting targets}
\State $N_k^p \gets M_p\cdot\hat{\nu}_k^p$ \Comment{Number of particles to resample by}
\State Resample $N_k^p$ times from $\cup_{i=1}^{N_{k-1}}(\mathbf{x}_{k|k-1,p}^{(i)},w_{k|k,p}^{(i)}/\hat{\nu}_k^p)$ to obtain $\cup_{i=1}^{N_{k-1}}(\mathbf{x}_{k,p}^{(i)},w_{k,p}^{(i)})$ where $w_{k,p}^{(i)} = \hat{\nu}_k^p/N_k^p$

\Ensure 
	\Statex Persistent PHD particle system $\cup_{i=1}^{N_{k-1}}(\mathbf{x}_{k,p}^{(i)},w_{k,p}^{(i)})$ after resampling
	\Statex Newborn PHD particle system $\cup_{i=1}^{N_{k}^b}(\mathbf{x}_{k|k-1,b}^{(i)},w_{k|k,b}^{(i)})$ after final update step
\end{algorithmic}
\hrulefill
\caption{Multi-sensor adaptive birth intensity predict/update step algorithm.}\label{alg:abi_phdfilt_multisensor}
\end{figure*}

\subsection{Multi-sensor Adaptive Birth Intensity PHD Filter}
Given the particle filter implementation of the single-sensor adaptive birth intensity PHD filter \cite{Ristic2012a}, our extension to multi-sensor based on the iterated-corrector rule is straight forward.
The pseudocode for the resulting multi-sensor adaptive birth intensity PHD filter is shown in Figure~\ref{alg:abi_phdfilt_multisensor}.
Each step of the filter operates on the combined persistent (after resampling) and newborn particle system from the previous step, using the measurements generated by each sensor $l \in {1,\ldots,L}$.
The predict step (lines 1-5) is similar to what is described in \cite{Ristic2012a}, and based on target survival and kinematic propagation.
Applying the iterated-corrector to the update step, the proposed approach samples a particle system using a birth intensity based on the measurements observed for a given sensor (lines 6-23).
The predicted persistent and newborn particle systems are updated with that sensor's observed measurements, denoted $\mathbf{Z}_k^{(l)}$, using the adaptive birth intensity PHD update steps defined by Equations~\eqref{eq:abi_phdpred_persupd} and \eqref{eq:abi_phdpred_birthupd} (lines 17-18).
Then, an intermediate ``predict" step occurs that converts all of the birth particles generated by that sensor (after performing the update of Equation~\eqref{eq:abi_phdpred_birthupd}) to persisting particles (line 19-22). The next sensor's birth particle system is sampled using its measurement set, and the update equations are repeated using the persistent particle set generated after the previous sensor's update.

It should be noted that the iterated-corrector multi-sensor combination rule is not commutative. As a result, sensors that are combined earlier in the update step will have a larger effect on the resulting persistent particle system than those combined later. For passive tracking applications, this could have the effect of introducing persistent ghost targets in the filter output  if sensors with higher false alarm rates are combined first. A common strategy used in passive radar applications is to randomize the combination order, favoring earlier selection of sensors with lower expected false alarm rates and higher detection rates. It should also be noted that the complexity of the proposed iterator-corrector update step grows with the number of measurements detected at each of the sensors, which can be very large for high false alarm rate sensors. However, this complexity growth is largely offset by the increase in particle placement efficiency provided by this technique over a uniform placement of particles. 


\section{Particle Sampling from Passive Measurements}\label{sec:sampling}
\begin{figure}
\centering
\includegraphics[width=0.48\textwidth]{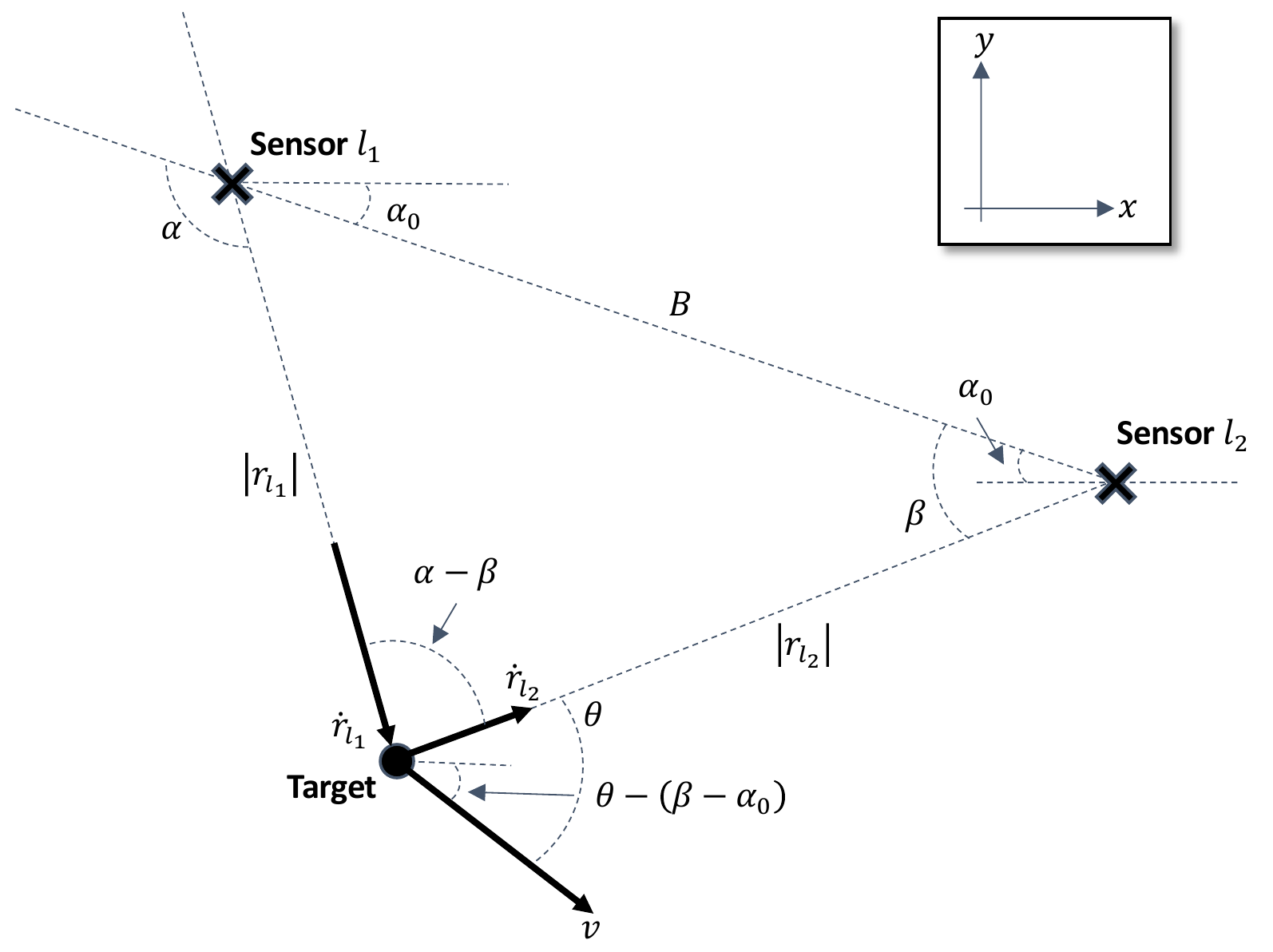}
\caption{TDOA/FDOA geometry diagram between two sensors and a target. Positive $x$ and $y$ coordinate system reference shown in upper right.}\label{fig:geometry}
\end{figure}

We now describe the application of the multi-sensor adaptive birth intensity filter to a passive multi-target target tracking problem using TDOA and FDOA measurements.
The TDOA and FDOA geometry for a single sensor pair and a single target is shown in Figure~\ref{fig:geometry}.
Let
$$
\mathbf{\Delta t}^{(l_1-l_2)}_{k} = \{\Delta t^{(l_1-l_2)}_{k,1},\ldots,\Delta t^{(l_1-l_2)}_{k,m^{(l_1-l_2)}_k}\}
$$
be the measurement RFS for TDOA measurements generated at time-step $k$ for sensor pair $(l_1,l_2)$. Similarly, let
$$
\mathbf{\Delta f}^{(l_1-l_2)}_{k} = \{\Delta f^{(l_1-l_2)}_{k,1},\ldots,\Delta f^{(l_1-l_2)}_{k,m^{(l_1-l_2)}_k}\}
$$
be the measurement RFS for FDOA measurements generated at time-step $k$ for sensor pair $(l_1,l_2)$.
The notation $(l_1-l_2)$ represents that the measurement difference is taken with respect to sensor $l_2$. We assume that the signal-level detection mechanism provides joint TDOA and FDOA detections that are independent and normally distributed for each true target such that
\begin{equation}\label{eq:tdoaSample}
\Delta t^{(l_1-l_2)}_{k,i} \sim \mathcal{N}\left(\frac{1}{c}(|r_{l_1}|-|r_{l_2}|),\sigma^2_{\Delta t}\right)
\end{equation}
and
\begin{equation}\label{eq:fdoaSample}
\Delta f^{(l_1-l_2)}_{k,i} \sim \mathcal{N}\left(\frac{f_c}{c}(|\dot{r}_{l_1}|-|\dot{r}_{l_2}|),\sigma^2_{\Delta f}\right)
\end{equation}
where $\sigma^2_{\Delta t}$ and $\sigma^2_{\Delta f}$ represent the expected TDOA and FDOA error variances, $c$ is the speed of light, and $f_c$ is the carrier frequency of the observation.
The quantities $|r_{l}|$ and $|\dot{r}_{l}|$ represent the range and range-rate from sensor $l$ to the target, where we have dropped the dependence on the location of the target for convenience.
Because the measurements are obtained jointly, their detections are characterized by a single TDOA/FDOA correct detection and false alarm rate probability per sensor. In the following sections, we describe a measurement driven birth particle sampling procedure using TDOA/FDOA measurements $$\mathbf{x}_{k|k-1,b}^{(n)} \sim b\left(\mathbf{x}_{k|k-1,b}^{(n)}|\left(\mathbf{\Delta t}^{(l_1-l_2)}_{k},\mathbf{\Delta f}^{(l_1-l_2)}_{k}\right)\right).$$
Without loss of generality, we consider the sampling technique for a single target at a fixed time step and for a single pair of sensors in the following subsections, denoting\begin{equation}\label{eq:rangerate}
\Delta r_{({l_1},{l_2})} = |r_{l_1}|-|r_{l_2}|
\end{equation}
and
\begin{equation}\label{eq:rangeratediff}
\Delta \dot{r}_{({l_1},{l_2})} = |\dot{r}_{l_1}|-|\dot{r}_{l_2}|
\end{equation}
for convenience.

\subsection{Birth Particle Sampling from TDOA Measurements}
The $x,y$ coordinates of the target assuming the coordinate system given in Figure~\ref{fig:geometry} is given as
\begin{equation}\label{eq:tdoa_pos}
\mathbf{x} = s_{l_1} - |r_{l_1}| \begin{bmatrix}
\cos(\alpha-\alpha_0) \\
\sin(\alpha-\alpha_0)
\end{bmatrix}
\end{equation} 
where the quantities $|r_{l_1}|$ and $\alpha$ must be sampled in order to be consistent with the observed TDOA value (\textit{i.e.}, the range difference $\Delta r_{({l_1},{l_2})}$). Applying law of cosines to Figure~\ref{fig:geometry}, we have that
\begin{align*}
|r_{l_2}|^2 &= |r_{l_1}|^2 + B^2 - 2B|r_{l_1}|\cos(\pi-\alpha), \\
& = |r_{l_1}|^2 + B^2 + 2B|r_{l_1}|\cos(\alpha).
\end{align*}
Noting that $|r_{l_2}|= |r_{l_1}|-\Delta r_{({l_1},{l_2})}$, the relationship between $|r_{l_1}|$ and $\alpha$ is given as \cite{Musicki2008}
\begin{equation}\label{eq:tdoa_r1alpha}
|r_{l_1}| = \frac{\Delta r_{({l_1},{l_2})}^2-B^2}{2(\Delta r_{({l_1},{l_2})} + B\cos(\alpha))}.
\end{equation} 
where $|r_{l_1}| \geq (\Delta r_{({l_1},{l_2})} + B)/2$. This lower bound is achieved when the target is on the line that intersects sensors $l_1$ and $l_2$ (\textit{i.e.}, $\alpha = \pi$).
In order to sample points uniformly along the hyperbola defined by Equations~\eqref{eq:tdoa_pos} and \eqref{eq:tdoa_r1alpha}, we first use the observed TDOA measurement to sample an estimate of the true range difference measurement, $\Delta r_{({l_1},{l_2})}$, according to the expected measurement variance and Gaussian assumption of Equation~\eqref{eq:tdoaSample}.
This sampled TDOA value is used to sample $|r_1|$ uniformly between $(\Delta r_{({l_1},{l_2})} + B)/2$ and some maximum value that is set based on the detection range of the sensor or the bounding box that defines the region of interest. Equation~\eqref{eq:tdoa_r1alpha} is then solved for $\alpha$ to yield
\begin{equation}\label{eq:tdoa_alphar1}
\alpha = \cos^{-1}\left( \frac{\Delta r_{({l_1},{l_2})}^2-B^2 - 2\Delta r_{({l_1},{l_2})}|r_{l_1}|}{2B|r_{l_1}|} \right)
\end{equation}
In this construction, the corresponding range on $\alpha$ covers the lower half of the hyperbola (\textit{i.e.}, positions beneath the line connecting the two sensors in Figure~\ref{fig:geometry}). To additionally consider the upper half of the hyperbola, we invert the sign of the inverse cosine argument in Equation~\eqref{eq:tdoa_alphar1} with probability $0.5$. The sampled values of $|r_1|$ and $\alpha$ are substituted into Equation~\eqref{eq:tdoa_pos} to generate the corresponding particle positions in the desired $x,y$ coordinate frame.


\subsection{Birth Particle Sampling from FDOA Measurements}
Given a particle location generated from a TDOA measurement, the angles $\alpha$ and $\beta$ can be determined uniquely by applying law of cosines. Using Figure~\ref{fig:geometry}, the observed range-rate difference of Equation~\eqref{eq:rangeratediff} for that particle is expressed as
\begin{align}
\Delta \dot{r}_{({l_1},{l_2})} &= |\dot{r}_{l_1}|-|\dot{r}_{l_2}| \nonumber\\
&= |v|\cos(\theta+(\alpha-\beta)) - |v|\cos(\theta) \nonumber\\
&= |v|\left[\cos(\theta+(\alpha-\beta)) - \cos(\theta)\right] \nonumber\\
&= -2|v|\sin\left(\frac{\alpha-\beta}{2}\right)\sin\left(\theta+\frac{\alpha-\beta}{2}\right)
\end{align}
where the final step results by noting that the linear combination of two cosines with arbitrary phases is a sinusoid with a scaled magnitude and phase shift. The resulting FDOA relationship between the velocity magnitude $|v|$ and the platform heading $\theta$ is then given as
\begin{equation} \label{eq:fdoa_vtheta}
|v| = -\frac{1}{2}\Delta \dot{r}_{({l_1},{l_2})}\left[ \sin\left(\frac{\alpha-\beta}{2}\right)\sin\left(\theta+\frac{\alpha-\beta}{2}\right)\right]^{-1}
\end{equation}
where 
\begin{equation}\label{eq:fdoa_vellb}
|v| \geq -\frac{1}{2}\Delta \dot{r}_{({l_1},{l_2})}\left[ \sin\left(\frac{\alpha-\beta}{2}\right)\right]^{-1}
\end{equation}
with equality when
$$
\theta+\frac{\alpha-\beta}{2} = \frac{\pi}{2}.
$$
In order to sample a velocity magnitude and bearing for each position particle generated by the TDOA measurements, we use the observed FDOA measurement to sample an estimate of the true range-rate difference, $\Delta \dot{r}_{(l_1,l_2)}$, according to the expected measurement variance and Gaussian assumption of Equation~\eqref{eq:fdoaSample}.
A velocity magnitude is uniformly sampled between the lower bound defined in Equation~\eqref{eq:fdoa_vellb} and some maximum possible velocity estimate. The corresponding heading of the target is then found by solving Equation~\eqref{eq:fdoa_vtheta} for $\theta$ to yield
\begin{equation}\label{eq:fdoa_thetav}
\theta = \sin^{-1}\left(\frac{-\Delta \dot{r}_{(l_1,l_2)}}{2|v|\sin\left(\frac{\alpha-\beta}{2}\right)}\right) - \frac{\alpha-\beta}{2}
\end{equation}
Similar to the TDOA case, we invert the sign of the inverse sine argument in Equation~\eqref{eq:fdoa_thetav} with probability $0.5$ in order to evenly samples headings with values greater than and less than $\frac{\alpha-\beta}{2}$.
The final velocity vector expressed in the $x,y$ coordinate system of Figure~\ref{fig:geometry} can then be fully determined as
\begin{equation}
\mathbf{v} = |v|\begin{bmatrix}
\;\;\;\cos(\theta-(\beta-\alpha_0)) \\ 
-\sin(\theta-(\beta-\alpha_0))
\end{bmatrix}
\end{equation}


\section{Tracking Simulation}\label{sec:simulation}
\subsection{Simulation Setup}\label{sec:setup}
\begin{figure}
\centering
\includegraphics[width=0.4\textwidth]{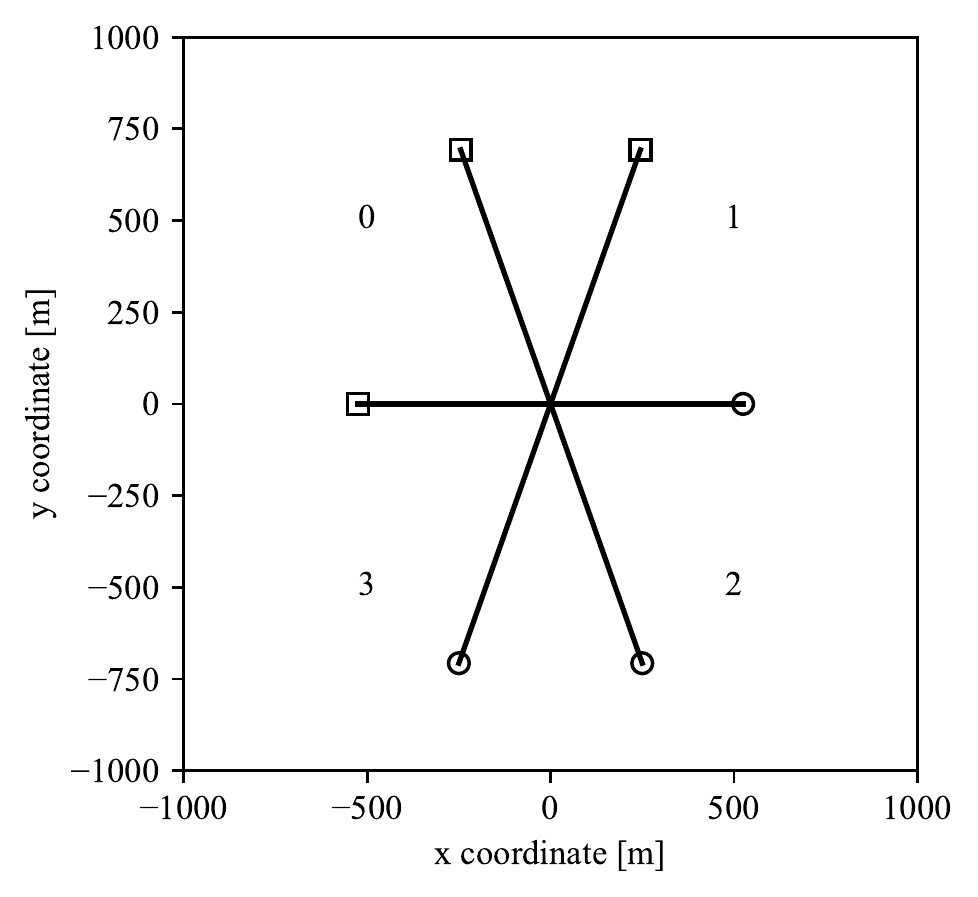}
\caption{Target trajectories shown in x-y plane.  Starting points denoted by $\circ$.  Stationary receivers shown as numbers.}\label{fig:laydown}
\end{figure}

In this section we numerically compare the tracking performance of the proposed TDOA/FDOA adaptive birth intensity PHD filter (denoted PHDF-M) versus a uniform birth intensity PHD filter of comparable particle density (denoted PHDF-U).
Figure~\ref{fig:laydown} shows the transmitter trajectories and receiver locations in a 2km $\times$ 2km area of interest. 
The velocity magnitude of all targets was fixed at 15 m/s.
The state vector of each target consisted of position and velocity in  $x$ and $y$ coordinates (\textit{i.e.}, $\mathbf{x} = [x, \dot{x}, y, \dot{y}]^\text{T}$).
A constant velocity transitional density, $\pi_{k|k-1}(\mathbf{x}|\mathbf{x'}) = \mathcal{N}(\mathbf{x}; \mathbf{Fx'}, \mathbf{Q})$, was assumed for persisting targets where
$$
\mathbf{F} = \mathbf{I}_2 \otimes \begin{bmatrix}
1 & T\\
0 & 1\end{bmatrix},
$$
$$
\mathbf{Q} = \mathbf{I}_2 \otimes q \begin{bmatrix}
T^3/3 & T^2/2\\
T^2/2 & T\end{bmatrix},
$$
$T$ is the sampling interval, $\otimes$ is the Kronecker product, $\mathbf{I}_m$ is the $m \times m$ identity matrix, and $q$ is the process noise intensity.
The simulation used $T = 1$ second and $q = 0.3$.
The probability of survival was set to $p_s(\mathbf{x}) = .98$.

\begin{figure}
\centering
\includegraphics[width=0.48\textwidth]{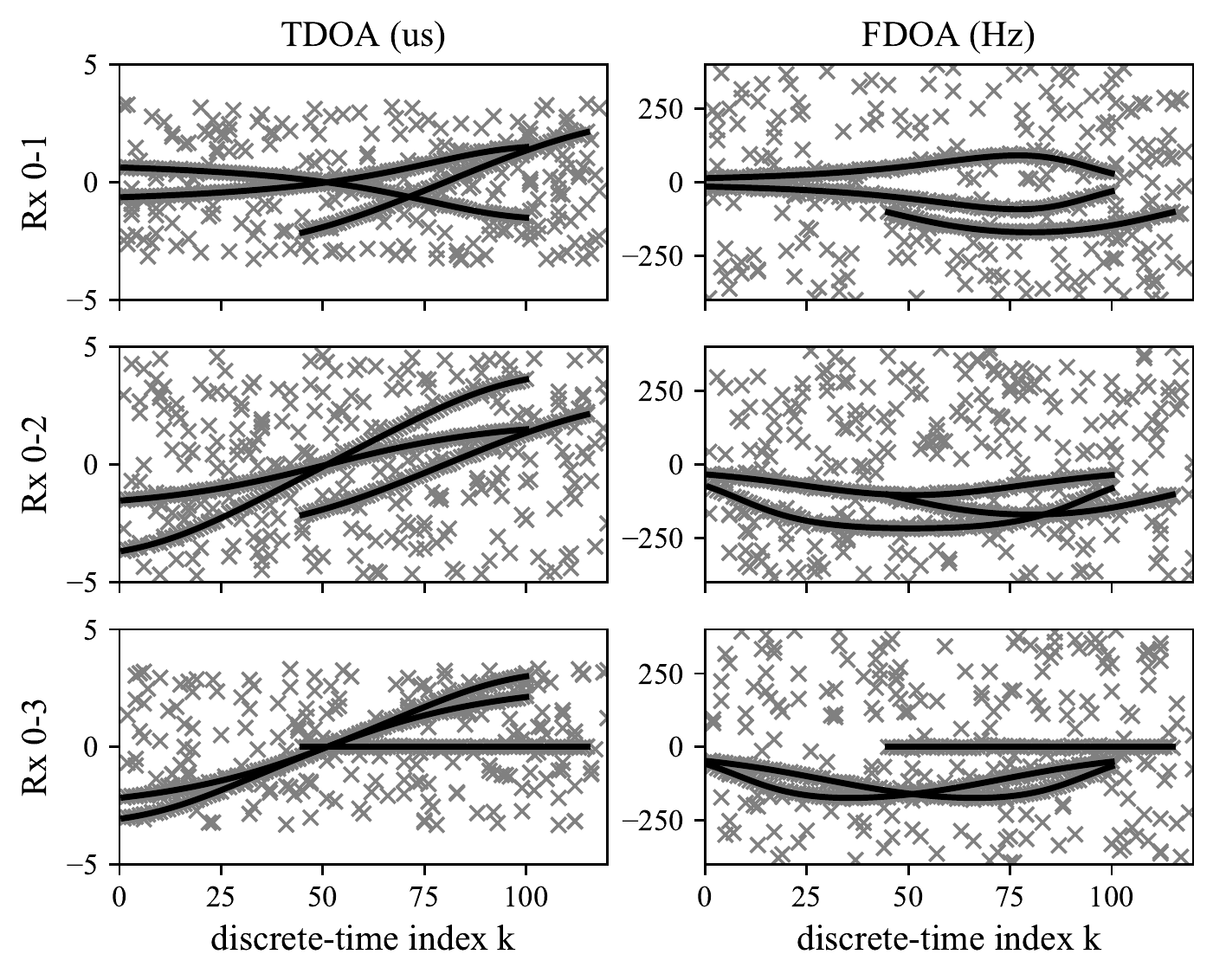}
\caption{Single realization of observed TDOA/FDOA measurements ($\times$) with respect to sensor 0 over time.  Measurements from actual targets shown as solid lines.}\label{fig:cluttermeas}
\end{figure}

TDOA and FDOA measurements were obtained from all sensor pair combinations $\{(0,1), (0,2), (0,3), (1,2), (1,3), (2,3)\}$.
As described in Equations~\eqref{eq:tdoaSample} and \eqref{eq:fdoaSample}, the measurement likelihood function for each sensor pair was set to $g_k(\mathbf{z}^{(l_1-l_2)} | \mathbf{x}) = \mathcal{N}(\mathbf{z}^{(l_1-l_2)}; \mathbf{h}^{(l_1-l_2)}(\mathbf{x}), \mathbf{R})$, where
\begin{equation}
\mathbf{h}^{(l_1 - l_2)}(\mathbf{x}) = \frac{1}{c} \begin{bmatrix}
(|r_{l_1}(\mathbf{x})|-|r_{l_2}(\mathbf{x})|)\\
f_c(|\dot{r}_{l_1}(\mathbf{x})|-|\dot{r}_{l_2}(\mathbf{x})|)
\end{bmatrix}
\end{equation}
and $\mathbf{R} = \text{diag}(\sigma^2_{\Delta t}, \sigma^2_{\Delta f})$.
The notations $|r_{l}(\mathbf{x})|$ and $|\dot{r}_{l}(\mathbf{x})|$ denote the range and range-rate from sensor $l$ to the target located in the state space at $\mathbf{x}$.
The timing and spectral measurements standard deviations were set as $\sigma_{\Delta t} = 20 \text{ns}$ and $\sigma_{\Delta f} = 2.5\text{Hz}$, respectively.
The clutter was uniformly distributed in time from $-cB^{(l_1-l_2)}$ to $cB^{(l_1-l_2)}$ and frequency from $-2|v| f_c / c$ to $2|v| f_c / c$, where $|v| = 25\text{m/s}$ and $f_c = 2.4\text{GHz}$.
Recall from the previous section that the quantity $B^{(l_1-l_2)}$ represents the straight-line distance between sensors $l_1$ and $l_2$.
The number of clutter points per receive window was Poisson distributed with a mean value of $\lambda = 2$.
For reference, a single realization of the clutter and true target measurements under this value of $\lambda$ is shown in Figure~\ref{fig:cluttermeas}.
The probability of detection was set as $p_d(\mathbf{x}) = 0.99$.
For every measurement $\mathbf{z}^{(l_1-l_2)} \in \mathbf{Z}^{(l_1-l_2)}_k$, newborn particles were birthed using the method described in Section~\ref{sec:sampling}, with $|r_{l_1}|\leq 2000\text{m}$, $|v| \leq 25\text{m/s}$, and $M_b = 500 \text{ particles/measurement}$.
To ensure a fair comparison between filters, the particle density of the PHDF-U was fixed to be the same as the PHDF-M density when all targets were present.
Finally, the  parameter $\nu_{k|k-1}^b = 0.0001$ was selected so that the average number of newborn targets per scan was $\hat{\nu}_k^b \approx 0.25$.

\subsection{Results}
\begin{figure}
\centering
\includegraphics[width=0.48\textwidth]{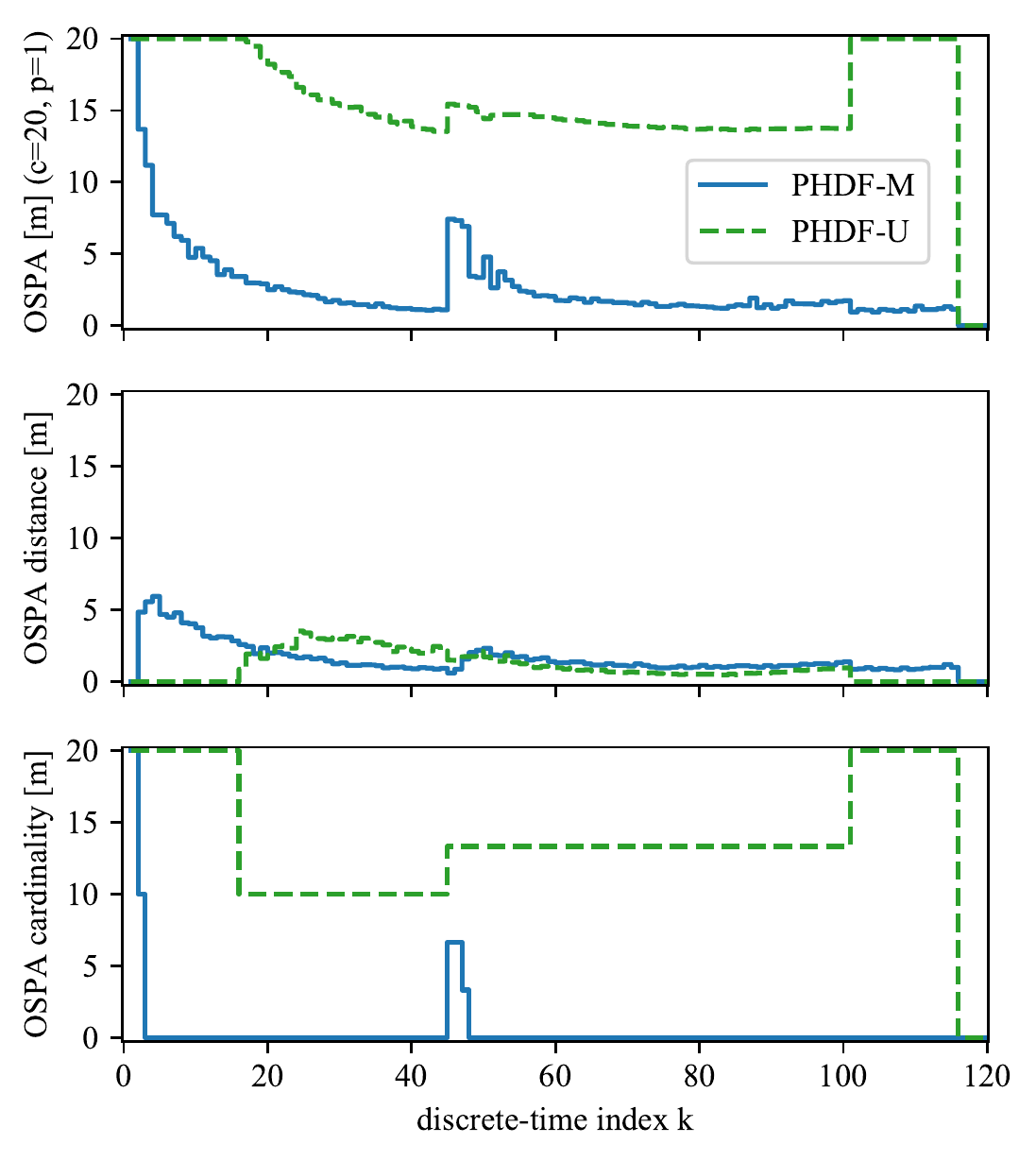}
\caption{100 MC run median OSPA metric (\textit{top}) over time for PHDF-M (\textit{solid blue line}) versus and PHDF-U (\textit{dashed green line});
median OSPA localization component (\textit{middle});
median OSPA cardinality component (\textit{bottom}).
}\label{fig:ospa}
\end{figure}

To capture median performance, $100$ Monte Carlo (MC) realizations of the tracks shown in Figure~\ref{fig:laydown} were simulated.
State extraction was performed using the method described in \cite{Ristic2010}, where the persistent particle system before resampling was used to estimate per particle measurement association weights and target existence probabilities.
The optimal subpattern assignment (OSPA) metric \cite{Schuhmacher2008b} was used to compare performance of the two filters.
Figure~\ref{fig:ospa} shows the median OSPA metric alongside the OSPA distance and cardinality components at each time step for the PHDF-M and PHDF-U filters. The OSPA cutoff parameter was set to $c = 20\text{m}$.
The PHDF-M cardinality performance significantly outperforms the PHDF-U. The PHDF-M was able to track all targets while the PHDF-U was only able to consistently track one of the three targets over the duration of the entire experiment.
For all tracked target states, the magnitude of the OSPA distance component of the PHDF-M and PHDF-U was similar.
To achieve equivalent OSPA cardinality and distance performance the number of particles of the PHDF-U had to be increased substantially.

\section{Conclusions and Future Work}
We have presented a technique for passive TDOA/FDOA tracking of multiple targets using the adaptive birth intensity PHD filter described in \cite{Ristic2012a}.
A multi-sensor extension of adaptive birth intensity PHD filter using the iterated-corrector technique was described alongside an exact technique for performing the state-space sampling of newborn particles directly from TDOA and FDOA measurements.
This technique was shown to adequately track the number and state of multiple targets with significantly less particles as compared to the uniform birth sampling technique typically used with the PHD filter.
Future work will involve the evaluation of the proposed tracking algorithm on passive measurements collected from a wireless sensor network of software defined radios.
Additionally, we will also be investigating sampling techniques for PDOA and extensions to the product multi-sensor fusion rule.

\bibliographystyle{IEEEtran}
\bibliography{IEEEabrv,SSL-TDOA-Fusion2018}

\end{document}